\documentclass[12pt]{article}
\usepackage{graphicx}
\usepackage{color}
\usepackage{amsmath}
\usepackage{units}
\usepackage{ifthen}
\usepackage[pdftex,bookmarks=true,bookmarksopen=true,bookmarksnumbered=true,a4paper=true,nesting=false,colorlinks=false]{hyperref}

\usepackage{wrapfig}
\usepackage{caption}
\usepackage{subcaption}
\usepackage{lineno}

\newcommand{\half}{\ensuremath{\frac{1}{2}}}

\newcommand{\ket}[1]{\ensuremath{ | #1 \rangle}}

\newcommand{\modOf}[1]{\ensuremath{\left| #1 \right|}}

\newcommand{\meas}[3]{\ensuremath{ #1 \pm #2_{\mathrm{stat}} \pm #3_{\mathrm{sys}}}}
\newcommand{\measNoTxt}[3]{\ensuremath{ #1 \pm #2 \pm #3}}
\newcommand{\lhcb}{{L}{H}{C}{b}}
\newcommand{\cleoc}{\mbox{{C}{L}{E}{O}{-}{c}}}
\newcommand{\besIII}{\mbox{{B}{E}{S}{ }{I}{I}{I}}}
\newcommand{\belleII}{\mbox{{B}{E}{L}{L}{E}{ }{I}{I}}}
\newcommand{\babar}{\mbox{{B}{a}{B}{a}{r}}}

\newcommand{\Araw}{\ensuremath{A_{\mathit{raw}}}}
\newcommand{\CP}{\ensuremath{\mathit{CP}}}
\newcommand{\CPV}{{\CP\ violation}}

\newcommand{\ppoint}{\ensuremath{\mathbf{p}}}

\newcommand{\Scp}{\ensuremath{S_{\CP}}}
\newcommand{\Scpi}{\ensuremath{S_{\CP}^i}}

\newcommand{\gam}{\ensuremath{\gamma}}
\newcommand{\prt}[1]{\ensuremath{\mathrm{#1}}}
\newcommand{\particle}[1]{\prt{#1}}
\newcommand{\NP}{New Physics}
\newcommand{\SM}{Standard Model}

\newcommand{\Bm}{\particle{B^{-}}}
\newcommand{\Bpm}{\particle{B^{\pm}}}

\newcommand{\Dee}{\prt{D}}
\newcommand{\Do}{\prt{D^0}}
\def\Dbar    {\kern 0.1em\overline{\kern -0.0em D}{}}

\newcommand{\Dobar}{\prt{\Dbar^0}}

\newcommand{\Done}{\prt{D^0_{1}}}
\newcommand{\Dtwo}{\prt{D^0_{2}}}

\newcommand{\Dstp}{\prt{D^{*+}}}
\newcommand{\Dstm}{\prt{D^{*-}}}

\newcommand{\Dp}{\prt{D^{+}}}

\newcommand{\Dpm}{\prt{D^{\pm}}}

\newcommand{\Dsp}{\prt{D_s^{+}}}

\newcommand{\Dspm}{\prt{D_s^{\pm}}}

\newcommand{\Kp}{\prt{K^{+}}}
\newcommand{\Km}{\prt{K^{-}}}
\newcommand{\Kpm}{\prt{K^{\pm}}}

\newcommand{\pip}{\prt{\pi^{+}}}
\newcommand{\pim}{\prt{\pi^{-}}}
\newcommand{\pipm}{\prt{\pi^{\pm}}}

\newcommand{\un}[2]{\ensuremath{#1\,\mathrm{#2}}}

\newcommand{\eqnref}[1]{Eq.~\ref{#1}}

\newcommand{\secref}[1]{Sec.~\ref{#1}}

\newcommand{\Figref}[1]{Figure~\ref{#1}}
\newcommand{\figref}[1]{Fig.~\ref{#1}}

\usepackage[sort&compress, sort, numbers]{natbib}
\usepackage{mciteplus}

\newboolean{articletitles}
\setboolean{articletitles}{true} % False removes titles in references

\newboolean{uprightparticles}
\setboolean{uprightparticles}{false} %True for upright particle symbols

\newboolean{inbibliography}
\setboolean{inbibliography}{false} %True once you enter the bibliography

\def\pbnr{}
\def\speaker{Jonas Rademacker}
\def\onbehalfof{}%the organisers}
\def\title{Measuring CP violation in 3- and 4-body decays}
\def\affiliation{H H Wills Physics Laboratory, University of Bristol, UK}
\def\support{}
\newcommand\pubnumber{\pbnr}
\newcommand\pubdate{\today}
\def\Title#1{\begin{center} {\Large #1 } \end{center}}
\def\Author#1{\begin{center}{ \sc #1} \end{center}}

\newcommand{\OnBehalf}[1]{\sbox0{#1}\ifdim\wd0=0pt
        {}% if #1 is empty
	\else
	{\\on behalf of #1}% if #1 is not empty
	\fi}
\newcommand{\SupportedBy}[1]{\sbox0{#1}\ifdim\wd0=0pt
        {}% if #1 is empty
	\else
	{\footnote{#1}}% if #1 is not empty
	\fi}
\def\Address#1{\begin{center}{ \it #1} \end{center}}

\newcommand\pubblock{\includegraphics[width=5cm]{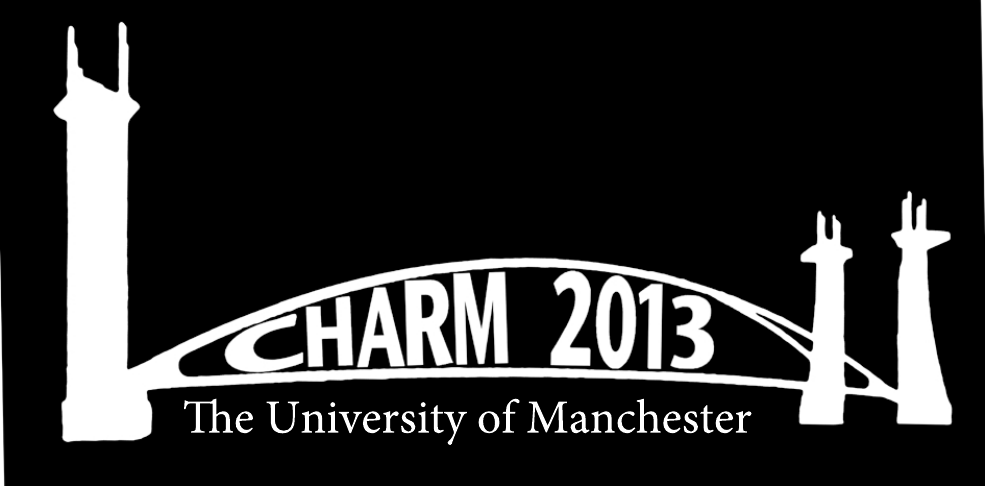}\hfill{\begin{tabular}{l} \pubnumber\\
         \pubdate  \end{tabular}}}
\newenvironment{Abstract}{\begin{quotation}  }{\end{quotation}}
\newenvironment{Presented}{\begin{quotation} \begin{center} 
             PRESENTED AT\end{center}\bigskip 
      \begin{center}\begin{large}}{\end{large}\end{center} \end{quotation}}
\def\Acknowledgements{\bigskip  \bigskip \begin{center} \begin{large}
             \bf ACKNOWLEDGEMENTS \end{large}\end{center}}
\def\venue{The 6$^{th}$ International Workshop on Charm Physics\\
(CHARM 2013)\\
Manchester, UK,  31 August -- 4 September, 2013}
\def\beq{\begin{equation}}
\def\eeq#1{\label{#1}\end{equation}}
\def\eeqn{\end{equation}}

\def\beqa{\begin{eqnarray}}
\def\eeqa#1{\label{#1}\end{eqnarray}}
\def\eeqan{\end{eqnarray}}

\let\bar=\overbar

\def\ket#1{\left| {#1} \right\rangle}

\def\half{\frac{1}{2}}

\def\Dslash{\not{\hbox{\kern-4pt $D$}}}
\def\dslash{\not{\hbox{\kern-2pt $\del$}}}

\def\msb{{\bar{\ssstyle M \kern -1pt S}}}

\newcommand*\patchAmsMathEnvironmentForLineno[1]{%
  \expandafter\let\csname old#1\expandafter\endcsname\csname #1\endcsname
  \expandafter\let\csname oldend#1\expandafter\endcsname\csname end#1\endcsname
  \renewenvironment{#1}%
     {\linenomath\csname old#1\endcsname}%
     {\csname oldend#1\endcsname\endlinenomath}}% 
\newcommand*\patchBothAmsMathEnvironmentsForLineno[1]{%
  \patchAmsMathEnvironmentForLineno{#1}%
  \patchAmsMathEnvironmentForLineno{#1*}}%
\AtBeginDocument{%
\patchBothAmsMathEnvironmentsForLineno{equation}%
\patchBothAmsMathEnvironmentsForLineno{align}%
\patchBothAmsMathEnvironmentsForLineno{flalign}%
\patchBothAmsMathEnvironmentsForLineno{alignat}%
\patchBothAmsMathEnvironmentsForLineno{gather}%
\patchBothAmsMathEnvironmentsForLineno{multline}%
}

\begin{document}
%%\linenumbers

\begin{titlepage}
\pubblock

\vfill
\Title{\title}
\vfill
\Author{\speaker\SupportedBy{\support}\OnBehalf{\onbehalfof}}
\Address{\affiliation}
\vfill
\begin{Abstract}
  Multibody charm decays have a rich phenomenology and potentially
  unique sensitivity to CP violation. In these proceedings we
  discuss recent results, challenges and prospects in searches for
  CP violation in three and four body charm decays.
%%%%%%%%%%%%%%%%%%%%%%%%%%%%%%%%%%%%%%%%%%%%%%%%%%%%%%%%%%%%%%%%%%%%%%%%%%%
% YOUR ABSTRACT GOES HERE
%%%%%%%%%%%%%%%%%%%%%%%%%%%%%%%%%%%%%%%%%%%%%%%%%%%%%%%%%%%%%%%%%%%%%%%%%%%
\end{Abstract}
\vfill
\begin{Presented}
\venue
\end{Presented}
\vfill
\end{titlepage}
\def\thefootnote{\fnsymbol{footnote}}
\setcounter{footnote}{0}
%

%%%%%%%%%%%%%%%%%%%%%%%%%%%%%%%%%%%%%%%%%%%%%%%%%%%%%%%%%%%%%%%%%%%%%%%%%%%
%  WHAT FOLLOWS IS YOUR TEXT
%%%%%%%%%%%%%%%%%%%%%%%%%%%%%%%%%%%%%%%%%%%%%%%%%%%%%%%%%%%%%%%%%%%%%%%%%%%
\section{Introduction}
In the search for physics beyond the \SM, the search for Charge-Parity
(\CP) violation 
%%(\CPV) 
in charm plays a special role. Charm provides a unique probe of \NP\
contributions involving flavour-changing neutral currents in up-type
quarks. The \SM\ prediction of zero or small \CPV\ in charm is rather
precise (although slightly less precise than most of us had thought
until a couple of years ago). The enormous, clean charm samples
available at current and future facilities promise unprecedented
precision in charm in a rich variety of decay channels.

\CPV, at least in the \SM, is due to \CP-violating weak phases. These are
observable when at least two amplitudes, $A_{1}$ and $A_{2}$, with different
strong and weak phases interfere. The resulting differential
decay rate to the phase-space point \ppoint\ is
\begin{equation}
 \frac{d\Gamma}{d\ppoint}(\ppoint) 
= \left[ 
  \modOf{A_1(\ppoint)}^2 +  \modOf{A_2(\ppoint)}^2 +  
  2\modOf{A_1(\ppoint)}\modOf{A_2(\ppoint)}
  \cos\!\left(\Delta \delta_s(\ppoint) + \phi_{\CP}(\ppoint)\right) 
\right] \frac{d\Phi}{d\ppoint},
\label{eq:A1A2}
\end{equation}
where $\frac{d\Phi}{d\ppoint}$ represents the density of states at
\ppoint, $\Delta \delta_s$ is the \CP-conserving strong interaction
phase difference between $A_1(\ppoint)$ and $A_2(\ppoint)$, while
$\phi_{CP}$ is the \CP-violating phase difference that changes sign
under \CP.
For two-body decays, $\frac{d\Phi}{d\ppoint}$ is a $\delta$ function
as only one (zero-dimensional) phase space point is
available. However, for $3, 4, 5, \ldots$ body decays, the allowed
phase space is at least $2, 5, 8, \ldots$ dimensional. The strong
phase difference varies across phase space, and so might the \CP\
violating phase. This results in a much richer phenomenology than for
two body decays, and motivates the search for local \CPV\ in
different phase space regions, while integrating over all phase space
might wash out any effects.

\section{Common Analysis Features}
\subsection{Choice of decay mode  - CF, SCS and DCS decays}
The sensitivity to \CPV\ is related to the relative size of the
interference term in \eqnref{eq:A1A2}, which is largest when $A_1$ and
$A_2$ are of similar magnitude. Significant direct \CPV\ is therefore
most likely in singly Cabibbo suppressed (SCS) decays, as here both
tree and penguin amplitudes contribute with comparable magnitude. In
Cabibbo favoured (CF) decays, the large SM tree contribution dominates
and significant \CPV\ effects are unlikely. For doubly Cabibbo
suppressed (DCS) decay amplitudes such as \prt{\Do \to
  K^+\pi^-}\footnote{The charge-conjugate mode is always implied
  unless stated otherwise}, there is no \SM\ penguin
contribution. However, here the amplitude proceeding via \prt{\Do \to
  \Dobar \to K^+\pi^-} is of a comparable magnitude; these ``wrong
sign'' (WS) decays are therefore sensitive to \CPV\ in \Dee\ mixing
and in the interference between mixing and decay.

\subsection{Tagging and the special role of the charm threshold}
At the B factories and at hadron colliders, the flavour of neutral D
mesons at production is identified by reconstructing \Dee\ mesons
resulting from the process \prt{\Dstp \to \Do \pi^+} and \prt{\Dstm
  \to \Dobar \pi^-}, where the charge of the slow pion identifies the
flavour of the \Dee.

At the charm threshold, where \cleoc\ and \besIII\ operate, \Dee\
mesons originate from the decay \prt{\psi(3770) \to \Dee \Dbar}. The
two \Dee\ mesons are quantum-correlated and have opposite flavour and
opposite \CP. This can be used to identify not only flavour
eigenstates, but a variety of well-defined \Do--\Dobar\
superpositions. For example when one \Dee\ (the tagging \Dee) decays
to a \CP-even eigenstate such as \prt{\Dee \to K^+ K^-}, the other
\Dee\ is identified as a \CP-odd superposition of \Do\ and
\Dobar. This is used to study the interference of \Do\ and \Dobar\
amplitudes in decays such as \prt{D \to K_S\pip\pim} or \prt{D \to
  \Kp\pim\pip\pim}~\cite{Libby:2010nu,Briere:2009aa,Insler:2012pm,
  Lowery:2009id,CLEO:DeltaKpi}, which is of considerable importance to
a variety of analyses, including \Dee\ mixing and
\CPV~\cite{Malde:2011mk,Bondar:CharmMixingCP} and the measurement of
the \CP\ violating phase $\gamma$~\cite{Atwood:coherenceFactor,GLW1,
  GLW2, ModelIndepGammaTheory, ADS, DalitzGamma1, DalitzGamma2,
  Rademacker:2006zx, BaBar_uses_us:2011up,LHCb2012DalitzGamma,
  LHCb-CONF-2013-006,LHCb2013GammaCombination,LHCb2013ADSObservation}. In
\secref{sec:ModelIndependence} we briefly discuss how mixing itself can be
used to improve the input from charm to the determination of
$\gamma$~\cite{MixingForGamma}.

\section{Time-integrated Analyses}
\subsection{Model-dependent Analyses}
Comparing results of amplitude fits to the phase space (Dalitz)
distribution of multibody decays for \CP-conjugate decay modes
provides a measure \CP\ violation. The advantage of this approach is
the physical interpretation of any \CPV\ observation that such a fit
result would allow. Recent searches using this approach include
\cleoc's\ amplitude analysis in \prt{\Do \to
  \Kp\Km\pip\pim}~\cite{cleoKKpipi} and CDF's analysis in \prt{\Do\
  \to K_S\pip\pim}~\cite{CDF_KSpipi}. In the CDF analysis, based on
$350\,000$ \prt{\Do \to K_S\pi\pi} events~\cite{CDF_KSpipi}, the
statistical uncertainty is already of a similar magnitude as the
systematic uncertainty, which is dominated by the model
uncertainty. This indicates that with the data samples that will be
available at LHCb and its upgrade, and at BELLE~II, the model-induced
systematic uncertainty would severely limit the precision of the
analyses. One option is to use model-independent methods, some of
which are described below. An alternative is to improve the
theoretical description of amplitude
models~\cite{Bigi:2013aca,Hanhart:2013afa}.

\subsection{$\chi^2$ based \CPV\ searches}
\label{sec:chi2CPV}
\subsubsection{Technique}
This class of model-independent searches for local \CPV\ in multibody
decays follows a statistical method first established by
BaBar~\cite{Aubert:2008yd} and developed further
in~\cite{Bediaga:2009tr,Bediaga:2012tm}. The \prt{\Dee \to f} and
\prt{\Dbar \to \bar{f}}\footnote{where here, \Dee\ indicates any \Dee\
  meson, charged or neutral, and $f$ indicates a generic final state;
  the bar indicates the \CP\ conjugate.} phase space is divided into
bins. For each pair of \CP-conjugate bins (with label $i$) a pull
variable
%\begin{equation}
$\Scpi = \frac{\Delta N_i}{\sigma(\Delta N_i)}$
%\label{eq:Scp}
%\end{equation}
is defined, where $\Delta N_i$ is the global-rate-adjusted difference
between the \prt{\Dee \to f} event yield in bin $i$, $N_i$, and
\prt{\Dbar \to \bar{f}} yield in the \CP\ conjugate bin,
$\bar{N}_i$. The difference is corrected for global rate asymmetries
by the factor $\alpha$, which is the ratio of the total event yields
$\alpha = \nicefrac{\sum N_i}{\sum \overline{N}_i}$: $ \Delta N_i = N_i -
\alpha \overline{N}_i $.  This removes any sensitivity to global rate
asymmetries, but also makes the analysis robust against global
production and detection asymmetries. The denominator in the
definition of \Scpi, $\sigma(\Delta N_i)$, is the
uncertainty on the numerator. With sufficiently large data samples,
the sum over all \Scpi\ follows a $\chi^2$ statistics. This
can be translated into a $p$-value, which represents the probability
to obtain $\chi^2 = \sum_i \left(\Scpi\right)^2$ as large as that
found in a given experiment, or larger, under the assumption of no
\CPV. In the absence of \CPV, the distribution of \Scpi\ is expected to
follow a Gaussian of width 1 and mean zero.

\begin{figure}
\newcommand{\CDFPicHeight}{0.222\textwidth}
\includegraphics[height=\CDFPicHeight]{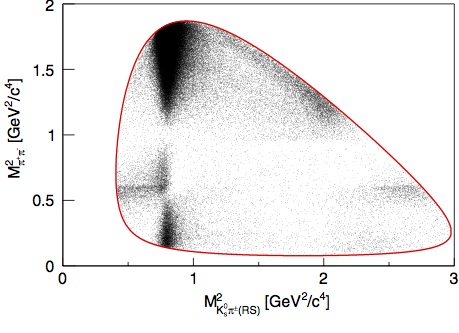}
\includegraphics[height=\CDFPicHeight]{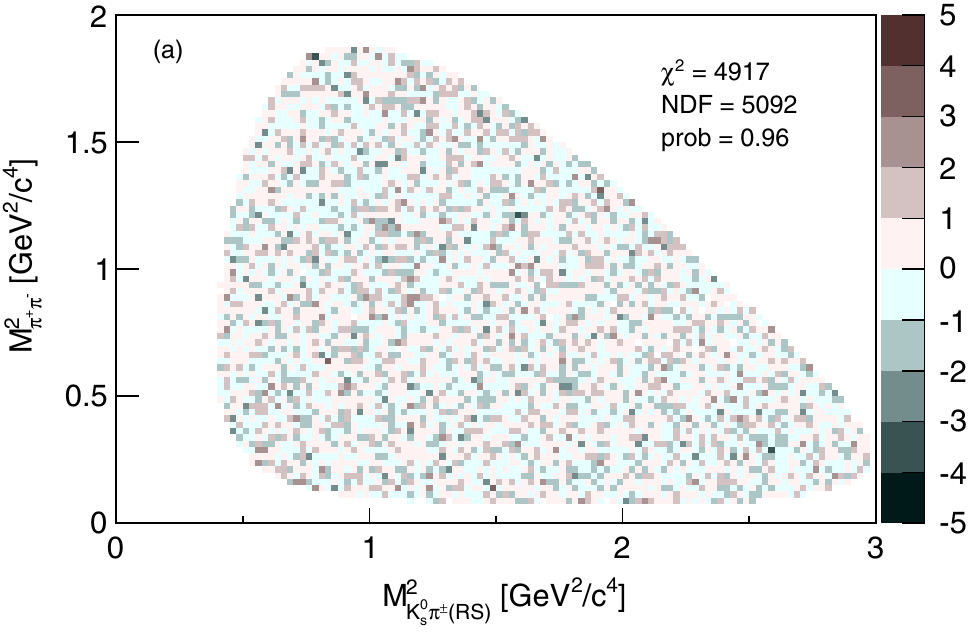}
\includegraphics[height=\CDFPicHeight]{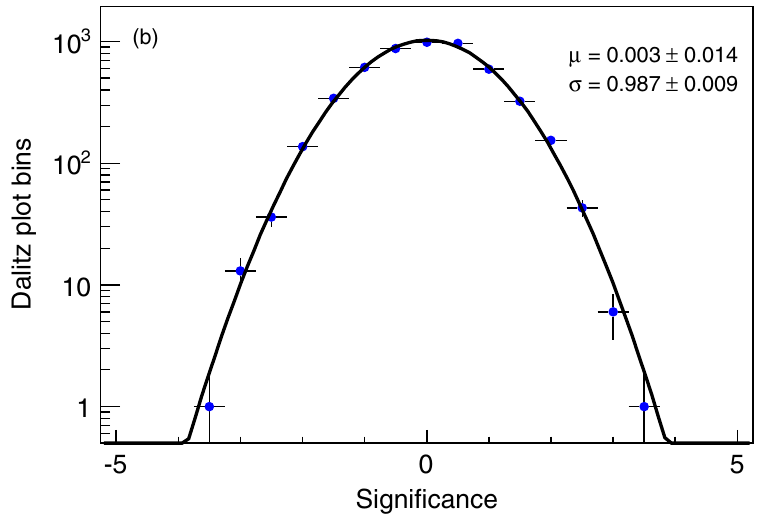}
\caption{From left to right: CDF's \prt{\Do \to K_S\pip\pim} Dalitz
  plot; the binning (colour code indicates $S_{CP}$ value); histogram
  of $S_{CP}$ values, with fit to a
  Gaussian~\cite{CDF_KSpipi}.\label{fig:CDF_KSpipi}}
\end{figure}
\subsubsection{Results for 3-body decays}
\babar\ published this year a search based on $223\,000$ \prt{\Dpm \to
  \Kp\Km\pipm} events in its full data sample.  LHCb analysed
$330\,000$ signal events in early data
(\un{35}{pb^{-1}})~\cite{LHCbDphipiI}, with recent updates in the
\prt{\Dpm \to \phi \pipm} region of the Dalitz
plot~\cite{LHCbDphipiII} based on a different technique discussed
below. CDF applied this method to $350\,000$ \prt{\Do \to
  K_S\pip\pim} events~\cite{CDF_KSpipi}; the Dalitz plot and \Scp\
distribution for this search are shown in
\figref{fig:CDF_KSpipi}. None of these analyses yielded evidence of
\CPV\ in charm.
\subsubsection{Results for 4-body decays}
The same principles used to search for \CPV\ in three-body charm
decays have been applied by the LHCb collaboration to four body SCS
decays~\cite{Matt}, with potentially unique sensitivity to CPV
effects~\cite{Bigi:2013aca}. The main difference to the three body
analyses is that the two-dimensional bins in the Dalitz plane are
replaced by five-dimensional hypercuboids. Analysing $57\,000$
\prt{\Do \to \Kp\Km\pip\pim} and $330\,000$ \prt{\Do \to
  \pip\pim\pip\pim} events, with very small background
contamination, no evidence for \CPV\ is found.
%%%

\subsection{Other binned methods}

\subsubsection{Phase-binning in \prt{\Dp \to \phi \pip}}
\begin{figure}
\begin{subfigure}{0.49\textwidth}
\includegraphics[width=0.99\textwidth]{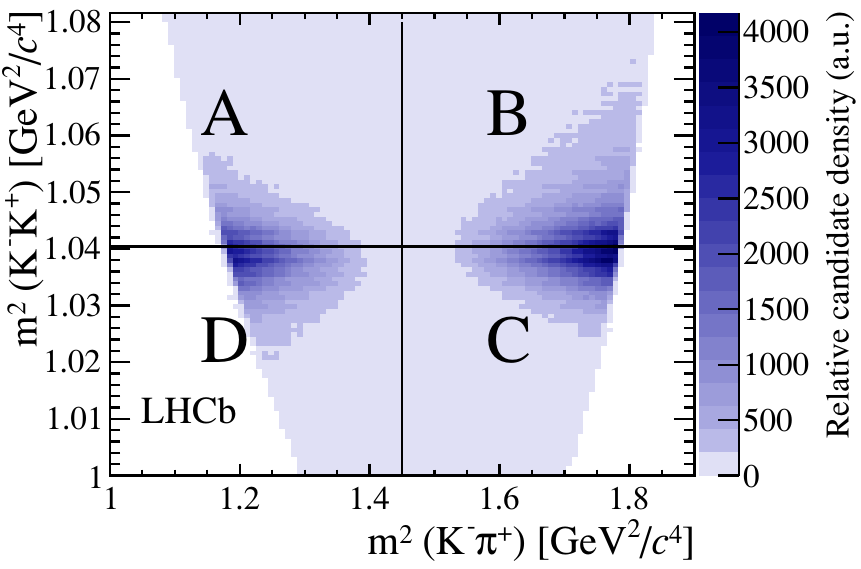}
\caption{\lhcb's \prt{D \to KK\pi} Dalitz plot in the $\phi$ region,
  split into four regions~\cite{LHCbDphipiII} based on the strong phase
  across the Dalitz plot. The asymmetries in each bin are used to
  construct the variable $A_{CP|S}$ described in the
  text.\label{fig:LHCbD2phipi}}
\end{subfigure}
\hspace{0.01\textwidth}
\begin{subfigure}{0.49\textwidth}
\centering
\includegraphics[width=0.75\textwidth]{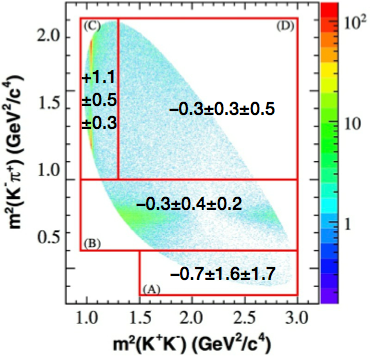}
\caption{BaBar's \prt{D \to KK\pi} Dalitz plot, split into four
  regions~\cite{BaBarD2KKpi}. The measured \CP\ asymmetries with
  statistical and systematic uncertainties are superimposed on the
  plot.\label{fig:BaBarD2KKpi}}
\end{subfigure}
\caption{Two different binning schemes based on the Dalitz structure
  of the decay.\label{fig:otherBinnings}}
\end{figure}
In a recent search for \CPV\ in the SCS decay \prt{D^+ \to
  \phi\pip}~\cite{LHCbDphipiII} at LHCb, the Dalitz plot around the
$\phi$ resonance is split into four regions, depending on the strong
phase $\delta$~\cite{Rubin:2008aa} (\figref{fig:LHCbD2phipi}). The
bins are chosen such that $\cos(\delta)$ is positive in regions $A$
and $C$, and negative in regions $B$ and $D$.  If \CPV\ is induced by
an amplitude with zero strong, and approximately constant
\CP-violating phase, the sign of $\cos(\delta)$ determines the sign of
the \CP-violating asymmetry. An observable optimised for this case is
constructed from the raw asymmetries \Araw\ in each region through $ A_{CP|S}
\equiv \half \left( \Araw^A + \Araw^C - \Araw^B - \Araw^D \right).  $
The raw asymmetries include production, detection and background
effects. To a good approximation, these effects cancel in
$A_{CP|S}$. The result, $A_{CP|S} = \left( \meas{-0.18}{0.17}{0.18}
\right)\%$, shows no sign of \CPV.
\subsubsection{Binning around resonances in \prt{\Dp \to \Kp\Km \pip}}
Several methods have been adopted by BaBar in a search for \CPV\ in
\prt{D^{\pm}} \prt{\to} \prt{\Kp\Km \pipm}~\cite{BaBarD2KKpi} based on
\un{476}{fb^{-1}} of data with approximately $223\,000$ selected
signal events. \Figref{fig:BaBarD2KKpi} illustrates a local \CPV\
search where the Dalitz plot is divided into bins around
resonances. This optimises the sensitivity e.g. if \CP\ violating new
physics affects one resonance and not the others. The measured
\CP-violating asymmetries in each bin are super-imposed on the Dalitz
plot, and show no evidence of \CPV.

\subsection{T-odd moments}
An alternative method for finding \CPV\ in four body decays such as
\prt{\Do} \prt{\to} \prt{\Kp\Km\pip\pim} is based on constructing $T$-odd
variables, i.e. variables that change sign under time-reversal
($T$)~\cite{BigiSanda}. For \Do\ and \Dobar\ decays respectively we
define
\begin{equation}
C_T = \vec{p}_{\Kp} \cdot 
\left( \vec{p}_{\pip} \times \vec{p}_{\pim} \right),
\;\;\;\;\;\;\;
\overline{C}_T = \vec{p}_{\Km} \cdot 
\left( \vec{p}_{\pim} \times \vec{p}_{\pip} \right).
\end{equation}
These variables are also parity ($P$)-odd, which is relevant in so far
as the $P$-reversed process with a sign change of all 3-momenta is directly accessible, while the $T$-reversed
process, resulting in the creation of a $D$ meson from
\prt{\Kp\Km\pip\pim}, is generally not.  The asymmetry $A_T$ between
$C_T > 0$ and $C_T < 0$, and its equivalent $\overline{A}_T$ for
\Dobar\ events,
\begin{equation}
A_T = \frac{\Gamma\left(C_T > 0\right) - \Gamma\left(C_T > 0\right)}{
      \Gamma\left(C_T > 0\right) + \Gamma\left(C_T > 0\right)},
\;\;\;\;\;
\overline{A}_T = \frac{\Gamma\left(-\overline{C}_T > 0\right) - \Gamma\left(-\overline{C}_T > 0\right)}{
      \Gamma\left(-\overline{C}_T > 0\right) + \Gamma\left(-\overline{C}_T > 0\right)},
\end{equation}
are parity violating observables, and $ \mathcal{A}_T \equiv
\half \left(A_T - \overline{A}_T\right) $ is \CP\ violating.  Searches
for \CP\ violation in this manner have been carried out by BaBar in
\prt{\Do \to \Kp\Km\pip\pim}, \prt{\Dp \to
  K^+K_S\pip\pim}, and \prt{\Dsp \to
  K^+K_S\pip\pim}~\cite{BaBarTodd1, BaBarTodd2}. No evidence for \CPV\ has been
found. It interesting to note that the values for $A_T$,
$\overline{A}_T$ vary considerably between the decay channels, with
larger values for \Do\ and \Dspm\ than for \Dpm:
\begin{align} 
\Do&: &             A_T &=\left(\measNoTxt{-68.5}{7.3}{5.8}\right)\cdot 10^{-3}
& \overline{A}_T &= \left(\measNoTxt{-70.5}{7.3}{3.9}\right)\cdot 10^{-3}
\notag\\
\Dspm&: &             A_T &=\left(\measNoTxt{-99.2}{10.7}{8.3}\right)\cdot 10^{-3}
& \overline{A}_T &=\left(\measNoTxt{-72.1}{10.9}{10.7}\right)\cdot 10^{-3}
\notag\\
\Dpm&: &             A_T &= \left(\measNoTxt{-11.2}{14.1}{5.7}\right)\cdot 10^{-3}
& \overline{A}_T &= \left(\measNoTxt{+35.1}{14.3}{7.2}\right)\cdot
10^{-3}
\notag
\end{align}
where the first uncertainty is statistical, and the second systematic.

\section{Time-dependent Amplitude Analyses}
\subsection{Formalism}
\Do\ and \Dobar\ mesons mix to form
mass eigenstates
%\begin{equation}
$
 \ket{\Done} 
 =  p \ket{\Do} + q |\Dobar \rangle
$ and $
 \ket{\Dtwo} 
 =  p \ket{\Do} - q | \Dobar \rangle
$
%\end{equation}
with masses $M_1, M_2$ and widths $\Gamma_1, \Gamma_2$. We define the
usual dimensionless parameters $x = 2\frac{M_1 - M_2}{\Gamma_1 +
  \Gamma_2}$ and $y = \frac{\Gamma_1 - \Gamma_2}{\Gamma_1 +
  \Gamma_2}$, where we follow the convention for the \CP\ operator
used by HFAG, where $\CP \ket{\Do} = - | \Dobar \rangle$, so that $D_1$ is
\CP\ even (see~\cite{MixingForGamma} for the impact of different
conventions on the definition of $x$ and $y$). The deviation of
$\modOf{\nicefrac{q}{p}}$ from unity parametrises \CPV\ in
mixing. \CPV\ in the interference between mixing and decay is
parametrised by the phase $\phi_D$; in the usual convention, this is
the phase of $\nicefrac{q}{p}$.

\subsection{Amplitude model-dependent analyses}
Time-dependent amplitude analyses in decays to final states that are
accessible to both \Do\ and \Dobar\ are sensitive to \CPV\ in mixing,
and \CPV\ in the interference between mixing and decay.  The latest
such and most precise measurement in \prt{\Do\ \to K_S\pi^+\pi^-} was
presented by Longke Li on behalf of the BELLE collaboration at this
conference~\cite{Li:2013mnl}. The results for the \CP-sensitive
parameters are
\[
\modOf{\nicefrac{q}{p}} =
  0.90^{+0.16}_{-0.15}\mbox{}^{+0.05}_{-0.04}\mbox{}^{+0.06}_{-0.05}
,\;\;\;\;\;
\phi_D = -6^{\circ}\pm 11^{\circ} \pm 3^{\circ} \mbox{}^{+3^{\circ}}_{-4^{\circ}}
\]
where the first uncertainty is statistical, the second systematic, and
the third represents the uncertainty due to the amplitude model
dependence. The results assume no direct \CPV. As in previous
measurements~\cite{CLEO_KspipiMix,BaBar_KspipiMix,BELLE_KspipiMix_1}
no evidence for \CPV\ was found. An example of a mixing and
\CPV-sensitive study using a non-self conjugate decay is that of
\babar's analysis of \prt{\Do \to
  K^+ \pi^- \pi^0}. \babar\ measured
$x^{\prime}_{\prt{K\pi\pi^0}}$ and $y^{\prime}_{\prt{K\pi\pi^0}}$
(related to $x$ and $y$ through a rotation in the $x-y$ plane)
separately for \Do\ and \Dobar,
\begin{align}
x^{\prime}_{\prt{K\pi\pi^0}} (\Do)
   &= \left(2.53^{+0.45}_{-0.63} \pm  0.39\right)\%
&
y^{\prime}_{\prt{K\pi\pi^0}} (\Do)
   &= \left(-0.05^{+0.63}_{-0.67} \pm  0.50\right)\%
\notag\\
x^{\prime}_{\prt{K\pi\pi^0}} (\Dobar)
   &= \left(3.55^{+0.73}_{-0.83} \pm  0.65\right)\%
&
y^{\prime}_{\prt{K\pi\pi^0}} (\Dobar)
   &= \left(-0.54^{+0.40}_{-1.16} \pm  0.41\right)\%,
\notag
\end{align}
also finding no evidence of \CPV.

\subsection{Model independence and the charm threshold}
\label{sec:ModelIndependence}
The theoretical uncertainty on the amplitude model in multibody
analyses potentially limits the precision that can be achieved in
measurements of mixing-induced \CPV\ at \lhcb\ and its upgrade, and at
\belleII. Model-independent methods tend to require input related to
the relative phases of the \Do\ and \Dobar\ decay amplitudes. This
information is accessible at the charm
threshold~\cite{Malde:2011mk,Bondar:CharmMixingCP,
  ChrisAndGuy2012}. This input is also important for measurements of the
\CPV\ parameter $\gamma$ in \prt{\Bpm \to D\Kpm} and related
decays~\cite{Libby:2010nu,Briere:2009aa,Insler:2012pm,
  Lowery:2009id,CLEO:DeltaKpi, Atwood:coherenceFactor,
  BaBar_uses_us:2011up,ModelIndepGammaTheory,LHCb2012DalitzGamma,
  LHCb-CONF-2013-006,LHCb2013GammaCombination,LHCb2013ADSObservation}.
The \cleoc\ collaboration provides such input for \prt{\Do \to
  K_S\pi\pi}, \prt{\Do \to K_S KK}~\cite{Libby:2010nu,Briere:2009aa},
\prt{\Do \to \Km\pip\pi^0}, \prt{\Do \to
  \Km\pip\pip\pim}~\cite{Lowery:2009id} (for a recent update see Guy
Wilkinson in these proceedings) and \prt{\Do \to K_S \Km\
  \Kp}~\cite{Insler:2012pm}.
Rather than seeing the dependence of charm mixing on charm coherence
parameters as an obstacle to precision mixing measurements, it can in
fact be used to constrain these coherence parameters and thus improve
the charm input to the measurement of \gam\ in \prt{\Bm \to DK^-} and
related decays. A recent study~\cite{MixingForGamma}, also discussed
in Samuel Harnew's contribution to these
proceedings~\cite{Harnew:2013mka}, indicates that with existing LHCb
data, input from charm mixing can substantially reduce the uncertainties
on the coherence parameters in \prt{\Do \to \Kp \pim\pip\pim}, with
the potential to significantly improve the precision of future
$\gamma$ measurements.

\section{Conclusion}
\CP\ violation in charm has unique sensitivity to physics beyond the
\SM. Multibody analyses offer a rich phenomenology and unique
sensitivity to \CPV, but also pose unique experimental and theoretical
challenges. In these proceedings, we give a summary of recent
experimental results, with impressive examples in terms the precision
achieved, and the complexity of the analyses required to achieve it.
In the face of rapidly increasing, high quality data samples, a
potential limitation for future analyses is the theoretical
uncertainty in the description of soft hadronic effects in multibody
decays. Model-independent methods, many relying on input from the
charm threshold, offer a way around this limitation in many cases. At
the same time, work is ongoing to reduce the theoretical uncertainty
in the description of multi-body decays. 
The very rich phenomenology accessible in multibody charm decays,
combined the prospect of large new data samples at LHCb (upgrade),
BELLE~II and at the charm threshold with BES~III, justify considerable
optimism for new and exciting results in multibody charm analyses in
the near future.

\Acknowledgements We thank the conference organisers. We acknowledge support from the Science and Technology
Research Council (UK) and the European Research Council under FP7.

\bibliographystyle{LHCb}
\bibliography{bibliography,bibliography2}

\end{document}